\documentclass[11pt,eqs]{article}

\usepackage{mathtools}
\usepackage{xcolor}
\usepackage[margin=1in]{geometry}
\usepackage{latexsym,amsmath,amssymb,amsthm,amstext,tikz}
\usepackage{fancyhdr}
\usepackage{graphicx,relsize}
\usepackage{multirow,multicol}
\usepackage{pdflscape}

\def\cleq{\setcounter{equation}{0}}

\DeclarePairedDelimiter\bra{\langle}{\rvert}
\DeclarePairedDelimiter\ket{\lvert}{\rangle}
\DeclarePairedDelimiterX\braket[2]{\langle}{\rangle}{#1 \delimsize\vert #2}
\DeclareMathOperator{\Tr}{Tr}

\title{Quantum theory phase space foundations\thanks{Work supported in part by The University of Belgrade, Belgrade, Serbia and by the home institutions of the authors.}}
\author{ Milo\v{s} D. Davidovi\'{c} \thanks{e-mail: davidovic@vinca.rs}, Ljubica D. Davidovi\'c \thanks{e-mail: ljubica@ipb.ac.rs}
and Milena D. Davidovi\'{c} \thanks{e-mail: milena@grf.bg.ac.rs}\\
{\it Institute for Nuclear Sciences Vin\v{c}a, Institute of Physics and }\\
{\it Faculty of Civil Engineering,}\\
{\it   University of Belgrade, Belgrade, Serbia}}

\begin{document}
\maketitle

\begin{abstract}

We give a review of concepts related to connection of classical and quantum theories, from the phase space perspective.
Quantum theory is described by non-commutative operators of coordinates and momenta, results in values having a certain probability,
hence resembling the classical statistical theory in which coordinates and momenta are well measurable quantities.
The search is for the classical object  best describing the quantum state, while the physical observables of the quantum theories are operators.

\end{abstract}

\section{Introduction}

The quantum mechanics defined on the phase space is a description that could lead us to the ultimate 
understanding of the connections between the classical and quantum mechanics, and the relations between the 
classical and quantum states of nature. Quantum mechanical states can be pure or mixed, a pure state cannot
become a mixed state without the interaction with it's environment and the real quantum systems are never isolated. To understand their relation to classical mechanics one should understand which of these states 
can behave classically.

The theoretical connection is made by
the canonical quantization, which transforms the classical phase space $(q,p)$ being a well measurable
property of a physical object
into the non-commuting operators $\hat{q},\hat{p}$ acting  on  physical states,
occupied only by a certain probability, which are not measurable simultaneously,
still knowing that the best results are obtained in the minimum of the uncertainty
relations.
The Weyl quantization precisely, is based on the 
Fourier transform between the phase space coordinates, together with its closure properties.

Often, one investigates the relation between 
the dynamics of physical processes and the phase space transformations, including the transformations of the functions defined on it,
in terms of the Weyl displacement operator
\begin{equation}
 D(q,p)=e^{-\frac{i}{\hbar}(q\hat{p}-p\hat{q})}=e^{\alpha \hat{a}^{+}-\alpha^\ast \hat{a}}=D(\alpha,\alpha^\ast).\notag
\end{equation}
This displacement of the system in the phase space plane,
which inevitably changes the density operator (describing the quantum state)
 was shown to be presentable in terms of identity resolution
(density) on phase space, named transition operator.
Reset to the initial density operator is done by the reconstruction operator. Interestingly enough, the Wigner function is the only quasi-distribution having the same transition and reconstruction operators \cite{AW}.

The exponentials of creation and annihilation operators, can in regard of the 
preferred ordering  yield different convergence properties. 
Normally ordered converge, anti-normal gain divergent complex integrals.
Also, representations on the ordinary or the complex phase space, have completely different diagonalization properties \cite{AW}. 
The first introduced non-negative quasi-distribution is the Husimi function, being the distribution for the simultaneous measurement for position and momentum operators with the maximal precision allowed by uncertainty relations. However, singling out the physical eigenvalues for Husimi is necessary. Husimi function is a convolution of Gaussian and Wigner functions \cite{LDB}.
The Wigner function itself is not always non-negative but smoothing it by an arbitrary function produces non-negativity. The choice of the smooth out function defines the class of non-negative distribution functions.
Problems of quasi-distributions include also completeness, over-completeness and normalization issues.

In reference \cite{MM} was shown that the non-validity of the normalization conditions for the concrete quantum states, described by quasi-probabilities given in terms of
 the Dirac delta function, the Airy function and the Fresnel integrals implies the corresponding density matrix can not be retrieved.

The conditions for a quantum state to behave classically were given in \cite{DL,DL1}.
It was proven $\lambda$-transformed states behave classically, by these conditions 
 and that they are necessarily mixed states \cite{DL2}. Therefore, the classically behaving states must interact with their surroundings at least in one part of their evolution.

In the phase space formulation,
quantum mechanical operators
are represented by their symbols, defined for the quasi-probability at hand, through an integral formula
for their expectation value
\begin{equation}
\langle\hat{O}\rangle=\int\!\!\!\int\,d\alpha d\beta\,
f(\hat{O})(\alpha,\beta)g(\hat{\rho})(\alpha,\beta).\notag
\end{equation}
The integration goes over the eigenvalues of the two chosen operators: coordinate and momentum, creation and annihilation operators
or two arbitrarily chosen operators. 
The Wigner function originates as the representation of the general quantum state in the real phase space,
while the  Husimi and  Glauber-Sudarshan  functions are connected to the coherent states, hence principally differing in  both foundation and  application.
Nevertheless
all three quasi-distributions can be represented by a $s$-parametized distribution $F(\alpha,s)$
on a complex phase space (characterized by $a,a^{+}$) \cite{JBM},
given by
\begin{equation}
F(\alpha,s)=\frac{1}{\pi^{2}}\int\,d^{2}\beta\,
G(\alpha,s)
e^{\alpha\beta^\ast-\alpha^\ast\beta},
\end{equation}
with $G$ being the generalized characteristic function
$
G(\alpha,s)=\Tr\Big{(}
\hat{D}(\beta)\hat{\rho}
\Big{)}e^{\frac{s}{2}|\beta|^{2}}.
$
For polynomials of creation and annihilation operators, symbols for these distributions correspond to symmetric, anti-normal and normal orderings, therefore it is believed that the linear interpolation of quasi-distributions bypasses the ordering problem \cite{AW}.
The connection is due to 
the quantum theory description of the harmonic oscillator dealing with
coordinate and momentum representations together with
states of the occupation number operator
and also coherent states on the same footing.
While creation and annihilation operators  count occupation numbers, they just re-scale the coherent states,
and these give Poissonian distribution of photons and Gaussian distribution of quadratures.
The examination of the exact statistics could reveal the border of the particle on the path description to the wave particle duality
including the wave interference.

The alternative to the operator formalism
is a deformation quantization,
where the formalism is replaced by a change of the multiplication between functions.
In quantum foundations scheme, this corresponds to the star product \cite{MS} acting between the operator symbols.

At the same time, there is also a large class of quasi-distributions defined by Cohen, which come from the resolution of the Dirac
delta product existing in the expectation value integrand
\begin{equation}
\langle\hat{O}\rangle=\int\!\!\!\int\!\!\!\int\,dxd\alpha d\alpha^\prime\,
\psi^\ast(\alpha)\psi(\alpha^\prime)\delta(\alpha-x)\hat{O}\delta(\alpha^\prime-x),\notag
\end{equation}
 and insertion of various identities for the eigenvector states of operators at hand.
So, 
the first class of quasi-distributions deals with explicit  values of $\psi^\ast(\alpha)\psi(\alpha^\prime)$, while the second class concentrates on representing $\delta(\alpha-x)\delta(\alpha^\prime-x)$.

It is interesting that in string theory, the representation of the delta function \cite{LD}
acting on two  string's space parameters causes the change in algebroidal properties,
including the form of the multiplication between the parameters labeling the generators of the allowed
background transformations.

\section{Wigner function}
\cleq

Wigner introduced the quasi-probability distribution function to describe the quantum corrections representing the fact that although the projectors to states with determined values of coordinate and momentum are known,
$\ket{q}\bra{q}$ and $\ket{p}\bra{p}$, the same is not true for the phase space coordinates $(q,p)$.
Following \cite{Manko.First}, one can obtain the Wigner function, comparing the coordinate and momentum representations of the density matrix.

For a discrete spectrum of a particle, a density matrix 
$\hat\rho=\sum_{n}w_{n}\ket{\varphi_{n}}\bra{\varphi_{n}}$ coordinate and momentum representation
are
\begin{equation}
\rho(q,q^\prime)=
\bra{q}
\hat\rho
\ket{q^\prime}=
\sum_{n}w_{n}\varphi_{n}(q)\varphi^\ast_{n}(q^\prime),
\end{equation}
\begin{equation}
\rho(p,p^\prime)=
\bra{p}
\hat\rho
\ket{p^\prime}=
\sum_{n}w_{n}\varphi_{n}(p)\varphi^\ast_{n}(p^\prime).
\end{equation}
Inserting  identities $\hat{1}_{q}=\int dq\ket{q}\bra{q}$,
into the diagonal momentum representation, one obtains
\begin{equation}
\rho(p,p)
=\frac{1}{(2\pi\hbar)^{1/2}}\int dqdq^\prime\,\rho(q,q^\prime)\,
e^{\frac{i}{\hbar}p(q^\prime-q)},
\end{equation}
with the plane wave
$\braket{q}{p}=\psi_{p}(q)=A e^{i\hbar^{-1}pq}$,
 being normalized to a Dirac delta function 
\begin{equation}
\psi_{p}(q)=\frac{1}{(2\pi\hbar)^{1/2}}e^{i\hbar^{-1}pq}.
\end{equation}
Then, re-scaling the coordinates, without losing the generality and coupling the Jacobiator to the 
coordinate not connected to the momentum, one obtains the definition of the Wigner function
\begin{equation}
W(p,q)
=\int du\,
\rho\big(q+\frac{u}{2},q-\frac{u}{2}\big)\,
e^{-i\hbar^{-1}pu}
\end{equation}
with a property
\begin{equation}
\rho(p,p)
=\int\,dq\, W(p,q),
\end{equation}
resembling the classical feature of the distribution function $f(p)=\int F(p,r)dr$.

\subsection{Wigner-Weyl symbols}

The first introduced quasi-distribution, the Wigner function is closely related to the Weyl quantization procedure.
Let us briefly recall the main features.
To the function on the phase space $A(q,p)$ expressed by an integral Fourier transformation as
\begin{equation}
A(q,p)=\frac{1}{(2\pi)^{2}}\int d\sigma\int d\tau\,\widetilde{A}(\sigma,\tau)e^{i(\sigma q+\tau p)},
\end{equation}
one associates the operator
\begin{equation}
\hat{A}(\hat{q},\hat{p})=
\frac{1}{(2\pi)^{2}}
\int d\sigma\int d\tau\,\widetilde{A}(\sigma,\tau)\,e^{i(\sigma \hat{q}+\tau \hat{p})},
\end{equation}
but additionally one should perform the symmetrization of the expansion, using the commutation relation, so that by 
Weyl correspondence to a classical  $q^{n}p^{m}$ one associates the symmetrical expression
\begin{equation}
 \frac{1}{2^{n}}\sum_{k=0}^{n}{n \choose k}\, \hat{q}^{n-k}\hat{p}^{m}\hat{q}^{k}=
\frac{1}{2^{m}}\sum_{l=0}^{m}{m \choose l}\, \hat{p}^{m-l}\hat{q}^{n}\hat{p}^{l}.
\end{equation}
The inverse Fourier transform closure, yields the appropriate Dirac delta representation.

When considering the complex phase space,
one can instead of
 the ordinary variables $a=(2\hbar)^{-1/2}\big(q+ip\big),a^{+}=(2\hbar)^{-1/2}\big(q-ip\big)$
consider 
\begin{eqnarray}\label{eq:aalam}
&&a=(2\hbar)^{-1/2}\big(\lambda q+i\lambda^{-1}p\big),
\nonumber\\
&&a^{+}=(2\hbar)^{-1/2}\big(\lambda q-i\lambda^{-1}p\big),
\end{eqnarray}
which still  obey $[{\hat a},{\hat a}^{+}]=1$. The Fourier transform of the function of the complex variable $A(\xi)\equiv A(\xi,\xi^\ast)$ is given by
\begin{equation}
\tilde{A}(\alpha)\equiv\tilde{A}(\alpha,\alpha^\ast)=\frac{1}{\pi}\int\,d^{2}\xi\,A(\xi,\xi^\ast)e^{a\xi^\ast-a^\ast\xi},
\end{equation}
with $d^{2}\xi=d{\Re}\xi\,d{\Im}\xi$,
and it follows 
using the Dirac delta function parity $\delta(x)=\delta(-x)$ that 
\begin{equation}
\delta^{(2)}(\xi)=\delta({\Re}\xi)\,\delta({\Im}\xi)=\frac{1}{\pi^{2}}\int\,d^{2}a\,e^{\,\xi a^\ast-\xi^\ast a}
=\frac{1}{\pi^{2}}\int\,d^{2}a\,e^{\,2{\Im}(\xi a^\ast)}.
\end{equation}
Introducing the displacement operator, an exponential function 
$\hat{D}(\alpha)=\exp\big(\alpha {\hat a}^{+}-\alpha^\ast {\hat a}\big)$,
one associates the operator $\hat{A}$ to the function $A(\xi)$  by
\begin{equation}
\hat{A}=\frac{1}{\pi}\int\, A(\xi)\hat{D}^{-1}(\xi)d^{2}\xi.
\end{equation}
The function is regained, using $\Tr\!\big[\hat{D}(\alpha)\hat{D}^{-1}(\beta)\big]=\pi \delta^{(2)}(\alpha-\beta)$, by
\begin{equation}
A(\alpha)=\Tr\!\big[\hat{A}\hat{D}(\alpha)\big].
\end{equation}
This is analogous to the characteristic function
introduced by Moyal in the real phase space
$C(\sigma,\tau)=\Tr\!\big{(}\hat\rho\hat{C}\big{)}=\Tr\!\big{(}\hat\rho\,
e^{\frac{i}{\hbar}(\sigma \hat{q}+\tau \hat{p})}\big{)}$, a Fourier transform of Wigner's quasi-distribution. 
In complex phase space,
the Wigner symbol corresponding to $\hat{X}$, is a Fourier transform of the
function $X$ reproducing $\hat{X}$ by the Weyl quantization rule
\begin{equation}
W_{\hat{X}}=\frac{1}{\pi}\int\,d^{2}\beta\Tr\Big{(}
\hat{X}\hat{D}(\beta)
\Big{)}e^{\alpha\beta^\ast-\alpha^\ast\beta}.
\end{equation}

\subsubsection{Wigner function as expectation value}
Let us find the characteristic function. Using the Backer-Hausdorf formula one can rewrite the displacement operator
in standard and antistandard form. In order to find the trace one inserts the identity resolutions  
$\hat{1}_{q}=\int dq\ket{q}\bra{q}$ and $\hat{1}_{p}=\int dp\ket{p}\bra{p}$, the first twice. Hence 
\begin{equation}
C(\tau,\sigma)=\frac{1}{2\pi\hbar}\int dp\int dq\int dq^\prime\,\sum_{k}w_{k}\varphi_{k}(q){\varphi}^{\ast}_{k}(q^\prime)e^{\frac{i}{\hbar}p(\tau+q^\prime-q)}
\begin{cases}
e^{\sigma(q-\frac{\tau}{2})},\quad\text{for}\; \hat{p}\hat{q}\\
e^{\sigma(q^\prime+\frac{\tau}{2})},\quad\text{for}\; \hat{q}\hat{p}.
\end{cases}
\end{equation}
Changing the coordinates into $q_\pm=\frac{1}{2}(q\pm q^\prime)$, and integrating over $p$ to obtain
$\delta(\tau-2q_{-})=\frac{1}{2}\delta(q_{-}-\frac{\tau}{2})$ one finds
\begin{equation}
C(\tau,\sigma)=\int dq_{+}\,\sum_{k}w_{k}\,
\varphi_{k}(q_{+}+\frac{\tau}{2})
{\varphi}^{\ast}_{k}(q_{+}-\frac{\tau}{2})\,
e^{\frac{i}{\hbar}\sigma q_{+}}.
\end{equation}
Its Fourier transformation is
\begin{equation}
C(\widetilde{q},\widetilde{p})=
\frac{1}{2\pi\hbar}\int d\tau\,\sum_{k}w_{k}\,
\varphi_{k}(\widetilde{q}+\frac{\tau}{2})
{\varphi}^{\ast}_{k}(\widetilde{q}-\frac{\tau}{2})\,
e^{-\frac{i}{\hbar}\tau\widetilde{p}}.
\end{equation}
In this way, the counterpart of ordering which is purely quantum physics technique vanished by use of coordinate re-scaling,
in obtaining the classical Wigner function, the Fourier transformed characteristic function.

Yet another method in calculating Wigner function can be found in \cite{SZ}
and it consists of square rooting  the translation generator while calculating in the coordinate representation,
followed by the cyclic permutation within trace.

\section{Husimi function}
\cleq

While the Wigner function is connecting coordinate and momentum representations of quantum state in the original interpretation,
or is connected to the expectation value of the parametrized displacement operator in the real phase space
(Weyl quantized characteristic function),
the connection of quasi-distributions $Q$ and $P$ to the classical theory is through coherent states.
The coherent states are the most classical quantum states of a harmonic oscillator,
and they can be generated by a classical oscillating current \cite{IQO}.
The modeling of quantum random walks is due to transition amplitudes between coherent states \cite{QprobQFT}.

The eigenvalues of position and momentum operators,
within Hamiltonian $\hat{H}=\frac{\hat{p}^{2}}{2m}+ U(\hat{x})$ where continuous, and their eigenstates orthogonal but normalized to Dirac delta function
\begin{equation}
\bra{x'}\hat{x}=\bra{x'}x',
\;
(x'-x)\braket{x'}{x}=0,\;
\braket{x'}{x}=\delta(x'-x),
\end{equation}
and the relation between coordinate and momentum representation of quantum state given by the Fourier transform
\begin{equation}
\Psi(p,t)=\frac{1}{\sqrt{2\pi\hbar}}\int\psi(q,t)e^{-iqp/\hbar}dq,
\end{equation}
along the path of the free particle with the assigned coordinate $q$.
But in quantum optics the harmonic oscillator with a quadratic potential $U(x)=\frac{1}{2}m\omega^2x^2$ became crucial in describing the motion of the trapped ions and atoms in standing waves cooled to temperatures 
revealing their specific quantum behavior, and also representing each mode  of the quantized electromagnetic field.
A single mode free field  described by the Hamiltonian $\hat{\cal H}=\hbar\omega(\hat{a}^{+}\hat{a}+\frac{1}{2})$.

The standard definition of coherent states is either through the annihilation or the displacement operator
\begin{equation}
\hat{a}\ket{\alpha}=\alpha\ket{\alpha},\quad \hat{D}(\alpha)\ket{0}=\ket{\alpha}.
\end{equation}
In general, the coherent states are the infinite sums of the number states, with  numbers representing the determined 
amount of the elementary quanta, which is up to a factor $\alpha$, labeling the states, left unchanged after the annihilation of an elementary quantum \cite{JPG}. The form of the expansion coefficients $\Phi_{n}(\alpha)$ at the coherent states, 
classifies them and determines the "optical phase space" as the image of the map
\begin{equation}
\alpha\rightarrow \xi_\alpha=\sqrt{\bar{n}(\alpha)}e^{i\arg{\alpha}},\quad
\bar{n}(\alpha)=\sum\,n|\Phi_{n}(\alpha)|^{2}.
\end{equation} 
The standard coherent states $\ket{\alpha}=e^{-|\alpha|^{2}/2}\sum_{n}\frac{\alpha^{n}}{\sqrt{n!}}\ket{n}$ are characterized by $\bar{n}(\alpha)=|\alpha|^{2}$. Ie. the coherent states are not represented by the exact number of quanta but with its distribution.

The Husimi function is defined for the standard coherent states of the harmonic oscillator,
with $\lambda=(m\omega)^{1/2}$ for its complex phase space variables (\ref{eq:aalam}), by
\begin{equation}\label{eq:H}
D_{H}(q,p)=\frac{1}{2\pi h}\bra{\alpha}\hat\rho\ket{\alpha}.
\end{equation}
The coherent state in the coordinate representation, in terms of the real phase space variables is $\braket{x}{\alpha}=(\frac{m\omega}{\pi\hbar})^{1/4}e^{-\frac{m\omega}{2\hbar}(x-q)^{2}+\frac{i}{\hbar}px}$.
Mathematically, what allows the definition of coherent states
is the representation of Heisenberg-Weyl group characterizing the classical phase space
on the space of square integrable complex functions, with group elements
\begin{equation}
e^{i\kappa\hat{1}+\frac{i}{\hbar}\big(
p\hat{Q}-x\hat{P}
\big)}.
\end{equation}
The 
general coherent basis is $\ket{q,p;\Phi}=\hat{U}(q,p)\ket{\Phi}$ with $\hat{U}(q,p)$ being the displacement operator
and the formula corresponding to  (\ref{eq:H}), defines
the smoothed Wigner function
\begin{equation}
D(q,p)=\frac{1}{2\pi}\bra{q,p;\Phi}\hat\rho\ket{q,p;\Phi},
\end{equation}
with the parameter $\Phi$ representing the fiducial vector, characterizing the quantization method \cite{QprobQFT}.
The discussion on smoothed Wigner function can be found in \cite{DL}.
The Husimi function is actually the Wigner function, smoothed by the "course graining" function
\begin{equation}
Q(q,p,\lambda)=\int d\tilde{q}\int d\tilde{p}\,W(\tilde{q},\tilde{p})\,
e^{-\lambda(\tilde{q}-q)^{2}-\frac{1}{\lambda}(\tilde{p}-p)^{2}}
\end{equation}
and it represents the pure squeezed state expectation value of the density matrix.
It is well known that the Husimi quasi-distribution \cite{ADM1,ADD1}, respond to the 
transformations of scale in the real phase space. So that,
under 
a scaling \cite{DL2} $(q,p)\rightarrow(\lambda q,\lambda p)$, the Husimi function $Q(q,p)$ of any
physical state is converted into a function which is also the Husimi function $\lambda^{2}Q(\lambda q,\lambda p)$ of a so called stretched physical state, provided that $|\lambda|^{2}\le 1$.

The coherent states have been generalized.
In use are the squeezed states, the states of radiation fields for which the fluctuation for one quadrature  is reduced from the symmetric quantum limit,
at the cost of enhanced second with no violation of Heisenberg uncertainty principle.
These are applicable in gravitational wave detection \cite{SZ},
because the amplitude of gravitational radiation is much smaller then the width of ground state wave function.
They can be generated by a classical oscillating current in presence of barriers or quadratic displacement potentials.

The squeeze operator, characterized by $\zeta=re^{i\theta}$, defined  by
\begin{equation}
\hat{S}(\zeta)=e^{\frac{1}{2}\big{(} \zeta^\ast\hat{a}^{2}-\zeta\hat{a}^{+2}\big{)}},
\end{equation}
 gives the squeezed coherent state by $\hat{D}(\alpha)\hat{S}(\zeta)\ket{0}=\ket{\alpha,\zeta}$.
Because the algebra of creation and annihilation operators is $[\hat{a},\hat{a}^{+}]=I$,
for their squares one has
\begin{eqnarray}
&&[\hat{a}^{2},\hat{a}^{+2}]=2\{\hat{a},\hat{a}^{+}\},\nonumber\\
&&[\hat{a}^{2},[\hat{a}^{2},\hat{a}^{+2}]]=8\hat{a}^{2},\nonumber\\
&&[\hat{a}^{+2},[\hat{a}^{2},\hat{a}^{+2}]]=-8\hat{a}^{+2},\nonumber\\
&&[\hat{a}^{+2},[\hat{a}^{2},[\hat{a}^{2},\hat{a}^{+2}]]]
=[\hat{a}^{2},[\hat{a}^{+2},[\hat{a}^{2},\hat{a}^{+2}]]]=-16\{\hat{a},\hat{a}^{+}\},\dots
\end{eqnarray}
and therefore needs the general form of the Baker-Cambell-Hausdorf relation in order to decompose the squeezing operator.
Note that $\{\hat{a},\hat{a}^{+}\}=\frac{2}{\hbar\omega}\hat{\cal H}$.
In matrix linear approximation, one can observe that three matrices closing this algebra share the same Lie algebra
\cite{BCH} with

$\frac{1}{4}\{\hat{a},\hat{a}^{+}\}\longleftrightarrow K_{0}=-\frac{1}{2}\sigma_{z}=\frac{1}{2}\begin{bmatrix}
-1 & 0\\
0 & 1
\end{bmatrix}$,
$\frac{1}{2}\hat{a}^{2}\longleftrightarrow 
K_{-}=\sigma_{+}=\begin{bmatrix}
0 & 1\\
0 & 0
\end{bmatrix}$ and 
$ \frac{1}{2}\hat{a}^{+2}\longleftrightarrow 
K_{+}=\sigma_{-}=\begin{bmatrix}
0 & 0\\
-1 & 0
\end{bmatrix}$,
and they obey 
$e^{cK_{0}}=\begin{bmatrix}
e^{-{c}/{2}} & 0\\
0 & e^{{c}/{2}}
\end{bmatrix}$,
$e^{cK_{-}}=I+cK_{-}=\begin{bmatrix}
1 & c\\
0 & 1
\end{bmatrix}$
and 
$e^{cK_{+}}=I+cK_{+}=\begin{bmatrix}
1 & 0\\
-c & 1
\end{bmatrix}$.

Therefore, the decomposition of the squeezing operator can be obtained by
equating the product of three matrices exponentials $e^{aK_{+}}e^{cK_{0}}e^{bK_{-}}$ and the representation of the squeezing operator 
$e^{-\zeta K_{+}+\zeta^\ast K_{-}}$
\begin{equation}
\resizebox{.9\hsize}{!}{$\exp\Big{\{}
\begin{bmatrix}
0 & \zeta^\ast\\
\zeta & 0
\end{bmatrix}
\Big{\}}=
\sum_{k=0}^\infty\frac{1}{(2k)!}\begin{bmatrix}
|\zeta|^{2k} &0\\
0 & |\zeta|^{2k}
\end{bmatrix}
+\sum_{k=0}^\infty\frac{1}{(2k+1)!}\begin{bmatrix}
0 & \zeta^\ast|\zeta|^{2k}\\
\zeta|\zeta|^{2k} & 0
\end{bmatrix}
=
\begin{bmatrix}
\cosh{|\zeta|} & \frac{\zeta^\ast}{|\zeta|}\sinh{|\zeta|}\\
\frac{\zeta}{|\zeta|}\sinh{|\zeta|} & \cosh{|\zeta|}
\end{bmatrix}.$}
\end{equation}
One finds
\begin{equation}
S(\zeta)=e^{-\frac{1}{2}e^{i\theta}\tanh{|\zeta|}\hat{a}^{+2}}
e^{-\ln(\cosh{|\zeta|})(\frac{1}{2}+\hat{a}^{+}\hat{a})}
e^{\frac{1}{2}e^{-i\theta}\tanh{|\zeta|}\hat{a}^{2}}.
\end{equation}
Hence, the squeezed vacuum state is given by
$\ket{0,0;\zeta}=(\cosh{|\zeta|})^{-1/2}e^{-\frac{1}{2}e^{i\theta}\tanh{|\zeta|}\hat{a}^{+2}}\ket{0}$.

There is also a non-unitary approach \cite{AWSS}, by which 
the squeezed vacuum states are
\begin{equation}
\ket{0,0;\zeta}\equiv(1+\zeta\zeta^\ast)^{1/4}\exp(-\frac{1}{2}\zeta a^{2+})\ket{0}
\end{equation}
 and they are eigenstates of the operator $a+\zeta a^{+}$ with the zero eigenvalue
$(a+\zeta a^{+})\ket{0,0;\zeta}=0,$ and normalized for $\braket{0,0;-\zeta}{0,0;\zeta}=1$.

\section{Glauber-Sudarshan function}
\cleq

The most used implicit definition of Glauber-Sudarshan quasi-distribution function,
is  the
quantization rule \cite{JPG} in terms of coherent states
\begin{equation}
\hat{\rho}=\frac{1}{\pi}\int\,d^{2}\beta\,P(\beta)\ket{\beta}\bra{\beta}.
\end{equation}
Obviously, the Husimi function is a  weighted   Glauber-Sudarshan quasi-probability,
because  
$|\braket{\beta}{\alpha}|^{2}=e^{-|\alpha-\beta|^{2}}$.
It is well known all three quasi-distributions $W,Q,P$ are connected by Weierstrass transform \cite{Polj}, smoothing out,
with $P$ being the  base of the smoothing  chain $Q\leftarrow W\leftarrow P$.

Mehta was first to determine the explicit form of the Glauber-Sudarshan quasi-distribution \cite{IQO},
by calculating
\begin{equation}
\bra{-\upsilon}\hat{\rho}\ket{\upsilon}=\frac{1}{\pi}e^{-|\upsilon|^{2}}\int\,d\alpha^{2}\,P(\alpha)e^{-|\alpha|^{2}}
e^{\alpha^\ast \upsilon-\alpha \upsilon^\ast},
\end{equation}
obtaining $P$-distribution as the weighted Fourier transform
\begin{equation}
\widetilde{P(\alpha)e^{-|\alpha|^{2}}}=e^{|\upsilon|^{2}}\bra{-\upsilon}\hat{\rho}\ket{\upsilon}.
\end{equation}
The optical equivalence theorem gives the expectation value of the normally ordered function by $\langle
G(a,a^{+})\rangle=\frac{1}{\pi}\int\,d^{2}\alpha P(\alpha)G(\alpha,\alpha^\ast)$.
Hence, there is a definition of the Glauber-Sudarshan P -- function as the expectation value of the 
normally ordered delta function
\begin{equation}
P(q,q^\prime)=\Tr\Big{(}
\hat{\rho}\delta(\alpha^\ast-\hat{a}^{+})
\delta(\alpha-\hat{a})
\Big{)},
\end{equation}
with
\begin{equation}
\delta(\alpha^\ast-\hat{a}^{+})\delta(\alpha-\hat{a})
=\frac{1}{\pi}
\int\,dc^{2}\,
e^{ic(\alpha^\ast-\hat{a}^{+})}
e^{ic^\ast(\alpha-\hat{a})}.
\end{equation}

\section{Cohen quasi-distributions}
\cleq

While calculating $W,Q,P$ distributions in different representations and relations between them,
one deals with differently ordered delta functions having operators within arguments, 
which lose the ordering after their application is done, possibly finding the new position 
according to the exact matching. These variations within the expectation value integral formula, are the main feature of the Cohen distributions.
The Cohen quasi-distributions are the result of combining properties of the transition amplitudes between the
eigenstates of the two operators chosen for the description. The main three being:
amplitudes are invertible, they close on 
Dirac delta and delta by definition satisfies $\int d\alpha\delta(\alpha-x)=1$. 

In paper \cite{BBC} authors start  by demonstrating the roots 
of ambiguities in representations of the  expectation value 
by considering the arbitrary operator,  $\langle g \rangle=\Tr[g\rho ]$, and operation of unity insertion
\begin{eqnarray}
\int dq_{1}\;\bra{q_{1}}g\rho\ket{q_{1}}&=&
\int dq_{1}\;\bra{q_{1}}\hat{1}_{p_{1}}g\hat{1}_{p_{2}}\hat{1}_{q_{2}}\rho\ket{q_{1}}
\nonumber\\
&=&\int dq_{1}\!\int dp_{1}\!\int dp_{2}\!\int dq_{2}\;\bra{p_{1}}g\ket{p_{2}}\braket{p_{2}}{q_{2}}\bra{q_{2}}\rho\ket{q_{1}}
\braket{q_{1}}{p_{1}}
\nonumber\\
&=&\int dp_{1}\;\bra{p_{1}}g\rho\ket{p_{1}}.
\end{eqnarray}
Within this formula one will straightforwardly obtain the Weyl-Wigner symbols for operators $g$ and $\rho$,
the Wigner function exactly for the second, by simple change of variables:
to difference and mean of both phase space variables, marked by $\bar{q},\bar{p}$ and $q,p$ respectively
\begin{eqnarray}
&&\langle g \rangle
=\int dq_{1}\!\int dp_{1}\!\int dp_{2}\!\int dq_{2}\;\bra{p_{1}}g\ket{p_{2}}\braket{p_{2}}{q_{2}}\bra{q_{2}}\rho\ket{q_{1}}
\braket{q_{1}}{p_{1}}
\\
&&=\int\int dqdp\;\frac{1}{2\pi\hbar}\int d\bar{p}\,\bra{p-\frac{\bar{p}}{2}}g\ket{p+\frac{\bar{p}}{2}}e^{-\frac{i}{\hbar}q\bar{p}}
\;\int d\bar{q}\,\bra{q+\frac{\bar{q}}{2}}\rho\ket{q-\frac{\bar{q}}{2}}e^{-\frac{i}{\hbar}p\bar{q}}.
\nonumber
\end{eqnarray}
Note that inserting unities in $p$-representation around $\rho$ in Wigner function $W(q,p)$,
it can 
for change of momentum $\bar{p}=p_{2}-p_{1}$, and after applying
$\frac{1}{2\pi\hbar}\int d\bar{q}e^{\frac{i}{\hbar}\bar{q}(p_{1}-p+\frac{\bar{p}}{2})}=\delta(p_{1}-p+\frac{\bar{p}}{2})$
become
\begin{equation}
W(q,p)=
\int d\bar{q}\,\bra{q+\frac{\bar{q}}{2}}\hat{1}_{p_{1}}\rho\hat{1}_{p_{2}}\ket{q-\frac{\bar{q}}{2}}e^{-\frac{i}{\hbar}p\bar{q}}
=\int d\bar{p}\,\bra{p-\frac{\bar{p}}{2}}\rho\ket{p+\frac{\bar{p}}{2}}e^{-\frac{i}{\hbar}q\bar{p}}.
\end{equation}
Hence 
\begin{equation}
\langle g\rangle=\int dq\int \frac{dp}{2\pi\hbar}\;W_{g}(q,p)W(q,p).
\end{equation}

Analogously, in the general case of two operators with given eigenvectors $\ket\alpha,\ket\beta$, values $\alpha,\beta$ and transition amplitudes $\braket{\alpha}{\beta}$, one can calculate the 
expectation value of the arbitrary operator by
\begin{equation}
\langle g \rangle=\int d\alpha^\prime\int d\alpha^{\prime\prime}{\cal A}^\ast(\alpha^\prime){\cal A}(\alpha^{\prime\prime})\delta(\alpha^{\prime}-\alpha^{\prime\prime})gd\alpha^{\prime}d\alpha^{\prime\prime}.
\end{equation}
Through delta and transition amplitudes one incorporates the other operator and again through delta and its closure one incorporates the second pair of
eigenvalues. The main interest for Cohen is the product of delta functions $\delta(\alpha-\bar\alpha)\delta(\beta-\bar\beta)$.
Considering their definition, by scaling the dumb variables and by insertions of its defining equalities (functions of $\alpha\beta$ and $\bar\alpha\bar\beta$ ) one can obtain various representations:
\begin{eqnarray}
&&\frac{1}{(2\pi)^{2}}\int d\sigma\!\int d\tau\, e^{i\sigma(\alpha-\bar\alpha)}e^{i\tau(\beta-\bar\beta)}
\nonumber\\
&=&\frac{1}{(\pi\hbar)^{2}}\int d\alpha^\prime\!\int d\beta^\prime
\, e^{-2i(\alpha^\prime-\alpha)(\beta^\prime-\beta)/\hbar}
e^{2i(\alpha^\prime-\bar\alpha)(\beta^\prime-\bar\beta)/\hbar}
\\
&=&\frac{1}{(2\pi)^{4}}\int d\alpha^\prime\!\int d\beta^\prime
\int d\theta\!\int d\theta^\prime\int d\tau\!\int d\tau^\prime
\frac{\Phi(\theta,\tau)e^{i\theta(\bar\alpha-\alpha^\prime)}e^{i\tau\theta\hbar/2}e^{i\tau(\bar\beta-\beta^\prime)}}{\Phi(\theta^\prime,\tau^\prime)e^{i\theta^\prime(\alpha-\alpha^\prime)}e^{i\tau^\prime\theta^\prime\hbar/2}e^{i\tau^\prime(\beta-\beta^\prime)}}\,.
\nonumber
\end{eqnarray}
The first will produce the Wigner quasi-distribution while the second gives the large class of Cohen quasi-distributions characterized by $\Phi$
\begin{equation}
\frac{1}{(2\pi)^{2}}\int d\alpha^\prime d\beta^\prime d\alpha^{\prime\prime}d\theta d\tau\,\Phi(\theta,\tau)
{\cal A}^\ast(\alpha^\prime){\cal A}(\alpha^{\prime\prime})
T(\bar\beta,\alpha^{\prime\prime})T^{+}(\alpha^{\prime},\bar\beta)
e^{i\theta(\bar\alpha-\alpha^\prime)}e^{i\tau\theta\hbar/2}e^{i\tau(\bar\beta-\beta^\prime)}.
\end{equation}

It is interesting, the insertion of the displacement operator unitary condition $DD^{+}=1$, together with its simple action on the creation and annihilation operators $D^{+}aD=a+\alpha$ and $D^{+}a^{+}D=a^{+}+\alpha^\ast$
induces displacement of states and "re-scaling" of operator
for example making transition from
  the squeezed coherent state to the squeezed vacuum expectation value
$\bra{\zeta,\alpha}a^{+k}a^{k}\ket{\xi,\alpha}\longrightarrow
\bra{\zeta}(a^{+}+\alpha^\ast)^{k}(a+\alpha)^{k}\ket{\xi}$ \cite{AA}.

\section{Electromagnetic field in the amplifier}
\cleq

Let us now consider a linear quantum amplifier model, described in the monograph \cite{MW}.
 We will consider a system of $N$ two-level atoms, $N_1$ of which are in the excited and $N_0$ in the ground state, with $N_0<N_1$,
kept constant in time,
interacting with  a single-mode quantum field.
The master equation for the electromagnetic field density operator is given,
in the Born-Markov approximation by the Lindblad equation
\cite{MW,AT}
\begin{eqnarray}   
\label{V3}
\frac {\partial\hat\rho}{\partial t}=-\gamma N_1(\hat a\hat a^{+}\hat\rho-2\hat a^{+}\hat \rho \hat a+\hat\rho \hat a\hat a^{+})-
\gamma N_0(\hat a^{+} \hat a\hat \rho-2\hat a\hat \rho \hat a^{+}+\hat \rho \hat a^{+} \hat a),
 \end{eqnarray}
with
$\gamma$ being the amplification coefficient. One can using the definition of the Husimi function, obtain the ordinary differential equation for the Husimi function,
and obtain the Husimi function for the amplified state  as in \cite{AT}:
 \begin{eqnarray}   
\label{V4}
Q(\alpha,t)=\frac{1}{\pi m}\int d^2 \beta\,Q(\beta)\,e^{-\frac{|\alpha-\beta G|^2}{m}},
 \end{eqnarray}
with $G(t)=e^{2(N_1-N_0)\gamma t}$ and $m=\frac{N_0}{N_1-N_0}(G^2-1)$.
 In \cite{ADM1,ADD1} we considered the case  where all atoms are in excited state so that $N_0=0$ and consequently $m=0$. Then, the expression for the Husimi function of the amplified state  is much simpler
\begin{eqnarray}   
\label{V6}
 Q_{out}(\alpha,t)=\frac1{G^2}Q_{in}(\alpha/G)=
\bra{\alpha/G}\hat\rho_{in}\ket{\alpha/G}.
\end{eqnarray}
Obviously, by considering the  dynamics of physical processes one encounters their relationship with both
  transformations of the phase space and the functions used to describe the process itself.
Depending on the occupation numbers of the content(media) which is here kept constant but generally fluctuating,
one deals with transformation of the phase-space being globally re-scaled or by over all phase-space function course graining.

 Let us now consider the bounded operator expansion in the normally, anti-normally and symmetrically ordered power series
\begin{equation}
F=\sum_{_{ N,M=0}}^\infty\, c_{_{ N,M}}\,(a^{+})^{N}a^{M}=
\sum_{_{ N,M=0}}^\infty\, d_{_{N,M}}\,a^{M}(a^{+})^{N}
=\sum_{_{ N,M=0}}^\infty\, e_{_{N,M}}\,\{a^{M}(a^{+})^{N}\},
\end{equation}
which are constrained by the existence \cite{CG} of
 the integral expansion 
\begin{equation}
F=\frac{1}{\pi}\int\Tr(FD(\xi))D^{-1}(\xi)d^{2}\xi.
\end{equation}
When this operator has passed the amplifier, its Husimi function becomes
\begin{eqnarray}
Q(\alpha,t)&=&\frac{1}{\pi m}\int d^2 \beta\,\sum_{_{ N,M=0}}^\infty\, c_{_{ N,M}}\beta^{\ast N}\beta^{M}
\,e^{-\frac{|\alpha-\beta G|^2}{m}},
\nonumber\\
Q(\alpha,t)&=&\frac{1}{\pi m}\int d^2 \beta\, \sum_{_{ N,M=0}}^\infty\,d_{_{ N,M}}\beta^{M}\beta^{\ast N}
\,e^{-\frac{|\alpha-\beta G|^2}{m}},
\nonumber\\
Q(\alpha,t)&=&\frac{1}{\pi m}\int d^2 \beta\,\sum_{_{ N,M=0}}^\infty\, \frac{e_{_{ N,M}}}{1+M}\sum_{n=0}^{M}\beta^{M-n}\beta^{\ast N}\beta^{n}
\,e^{-\frac{|\alpha-\beta G|^2}{m}}.
\end{eqnarray}
Hence, let us calculate the integral $I_{_{ N,M}}=\int d^2 \beta\,\beta^{M}\beta^{\ast N}
\,e^{-\frac{|\alpha-\beta G|^2}{m}}$.
Taking the change of variables $z=\frac{\alpha-\beta G}{\sqrt{m}}$, so that $\beta=\frac{\alpha-\sqrt{m}z}{G}$
one obtains
\begin{eqnarray}
I_{_{ N,M}}&=&\int d^{2}z\,\Big{(}-\frac{\sqrt{m}}{G}\Big{)}^{2}
\Big{(}
\frac{\alpha-\sqrt{m}z}{G}
\Big{)}^{M}
\Big{(}
\frac{\alpha-\sqrt{m}z}{G}
\Big{)}^{\ast N}
\,e^{-{|z|^2}}
\nonumber\\
&=&\int d^{2}z\,\Big{(}-\frac{\sqrt{m}}{G}\Big{)}^{2}\frac{1}{G^{^{N+M}}}
\sum_{i=0}^{M}{M\choose i}\alpha^{i}z^{M-i}
\sum_{j=0}^{N}{N\choose j}\alpha^{\ast j}(z^{\ast})^{N-j}
\nonumber\\
&\cdot&(-\sqrt{m})^{M-i+N-j}\,e^{-{|z|^2}}.
\end{eqnarray}
Transferring to the polar coordinates, it becomes
\begin{eqnarray}
I_{_{ N,M}}&=&
\frac{m}{G^{M+N+2}}
\int_{0}^\infty \rho d\rho \int_{0}^{2\pi}d\varphi\,
\sum_{i=0}^{M}{M\choose i}\alpha^{i}
\sum_{j=0}^{N}{N\choose j}\alpha^{\ast j}
\nonumber\\
&\cdot&
\rho^{M-i+N-j} e^{i\varphi(M-i-(N-j))}
(-\sqrt{m})^{M-i+N-j}\,e^{-{|z|^2}}.
\end{eqnarray}
The integration over $\varphi$ yields $\int d\varphi\,e^{i\varphi(M-i-(N-j))}=0$ for $M-i\neq N-j$. Hence
\begin{eqnarray}\label{eq:IMN}
I_{_{ N,M}}&=&
\frac{m}{G^{M+N+2}}
\int_{0}^\infty \rho d\rho 2\pi\,
\sum_{j=max(0,N-M)}^{N}{M\choose M-N+j}\alpha^{M-N+j}{N\choose j}\alpha^{\ast j}
\nonumber\\
&\cdot&
\rho^{2(N-j)} 
m^{N-j}\,e^{-{\rho^2}}.
\end{eqnarray}
Let us now consider the integration over $\rho$, marking
\begin{equation}
{\cal I}_{2(N-j)+1}=\int_{0}^\infty \rho d\rho\,\rho^{2(N-j)}\,e^{-{\rho^2}}=
\int_{0}^\infty d\rho\,\rho^{2(N-j)+1}\,e^{-{\rho^2}}.
\end{equation} 
The first integral is just
\begin{equation}
{\cal I}_{1}=\int_{0}^\infty d\rho\,\rho^{1}\,e^{-{\rho^2}}=
-\frac{1}{2}\int_{0}^\infty d(-\rho^{2})\,e^{-{\rho^2}}=-\frac{1}{2}e^{-{\rho^2}}\Big{|}_{0}^\infty=1/2.
\end{equation}
There is a recurrent relation, between consecutive integrals, obtained by the partial integration
\begin{equation}
{\cal I}_{2n+1}=\int_{0}^\infty d\rho\,\rho^{2n+1}\,e^{-{\rho^2}}=
-\frac{1}{2}\rho^{2n}e^{-{\rho^2}}\Big{|}_{0}^\infty
+n\int_{0}^\infty d\rho\,\rho^{2n-1}\,e^{-{\rho^2}}=n{\cal I}_{2n-1}.
\end{equation}
Therefore
${\cal I}_{3}=1{\cal I}_{1}=1/2$, ${\cal I}_{5}=2{\cal I}_{3}=1$, ${\cal I}_{7}=3{\cal I}_{1}=3$,
${\cal I}_{9}=4{\cal I}_{7}=4\cdot 3=12$ so that one can conclude
${\cal I}_{2n+1}=n!/2$ and confirm that
\begin{equation}
{\cal I}_{2n+3}=(n+1){\cal I}_{2n+1}=(n+1)n!/2=(n+1)!/2.
\end{equation}
In equation (\ref{eq:IMN}) we therefore substitute ${\cal I}_{2(N-j)+1}=(N-j)!/2$, and conclude
\begin{eqnarray}\label{eq:IMNfin}
I_{_{ N,M}}&=&
\frac{2\pi m}{G^{M+N+2}}
\sum_{j=max(0,N-M)}^{N}\frac{(N-j)!}{2}{M\choose M-N+j}\alpha^{M-N+j}{N\choose j}\alpha^{\ast j}
m^{N-j}
\nonumber\\
&=&\frac{\pi m}{G^{M+N+2}}
\sum_{j=max(0,N-M)}^{N}\frac{M!{N\choose j}}{(M-N+j)!}\alpha^{M-N+j}\alpha^{\ast j}
m^{N-j}.
\end{eqnarray}

The obtained functions are the benchmark used for evaluation of the actual statistics.

\section{Conclusion}

The connection between quantum and classical theories is established 
from theoretical perspective by the correspondence rules and from the 
physical by the quantum states of light probing the media, revealing their mutual statistics.
The correspondence rules in terms of quasi-distributions, are obtained once the integral representation of the operator expectation value is set.
In this paper, we found the Husimi quasi-probability function for the 
generically ordered operators fulfilling the evolution equation for the two level amplifier.
This example, in its settings, provides the explanation how both descriptions bundle.
Media induce
the  transformations of the phase space and the functions defined on it,
closely related to displacements and re-scaling,
transformations
which could also be achieved by a handful of theoretical treatments here revisited. 
Mainly, including the identity insertions, valid for the variables, transformations and operators at hand.



\begin{thebibliography}{}

\bibitem{AW} A. Wunsche, {\it The complete Gaussian class of quasi-probabilities and its relation to squeezed states and their discrete excitations,}
Quantum Semiclass. Opt. {\bf 8} (1996) 343-379.

\bibitem{LDB} D. Lalovi\'c, M.D. Davidovi\'c and N. Bijedi\'c, {\it Quantum mechanics in terms of non-negative smoothed Wigner function},
Physical Review A Vo. {\bf 46}, No. 3 (1992) 1206-1212.

\bibitem{MM}  V.I. Man’ko, L.A. Markovich, {\it Unnormalized quasi-distributions and tomograms
 of quantum states},
 Teoreticheskaya i Matematicheskaya Fizika, Vo. {\bf 197}, No. 2 (2018) 328–342.

\bibitem{DL}
M.D. Davidovi\'c and D. Lalovi\'c, {\it When does a given function in phase space belong to the class of Husimi distributions?},
J. Phys. A. Math. Gen. {\bf 26} (1993) 5099-5105.

\bibitem{DL1} D. M. Davidovi\'c and D. Lalovi\'c, {\it Quantum states that behave classically}, J. Phys. A:Math. Gen. {\bf 31} (1998) 2281-2285.

\bibitem{DL2} D. M. Davidovi\'c and D. Lalovi\'c, {\it The relation between the scaling of Husimi functions and the linear phase insensitive amplification of the corresponding quantum states and its implications}, J. Phys. A:Math. Gen. {\bf 29} (1996) 3787.

\bibitem{JBM} Jasel Berra–Montiel, {\it Star product representation of coherent state path
integrals},
Eur. Phys. J. Plus {\bf 135} (2020) 906.

\bibitem{MS} M. A. Soloviev, {\it Integral representations of the star product corresponding to
	the s-ordering of the creation and annihilation operators}, 
Phys. Scr. {\bf 90:7} (2015) 074008.

\bibitem{LD} Lj. Davidovi\'c, {\it Brackets in bosonic string theory}, J. High Energ. Phys. 2025, 177 (2025).

\bibitem{Manko.First} Yu. M. Belousov, V.I. Manko, {\it Density matrix. First part}
(MFTI, Moscow, 2004).

\bibitem{SZ} Marlan O. Scally and M. Suhail Zubairy, {\it Quantum optics}, (Cambridge University Press, Cambridge, 1997).

\bibitem{IQO} C. C. Gerry and P. L. Knight, {\it Introductory Quantum Optics},
(Cambridge University Press, Cambridge, 2004).

\bibitem{QprobQFT}
Aleksandra Pedrak, Andrzej Gózdz, Włodzimierz Piechocki, Patryk Mach, Adam Cieslik,
{\it Integral quantization based on the Heisenberg–Weyl group}, EPJ{\bf C 85} (2025) 617.

\bibitem{JPG} J.P. Gazeau, {\it Coherent States in Quantum Optics: an Overview},
chapter in Integrability, Supersymmetry and Coherent States (Springer, 2019) 69-101.

\bibitem{ADM1}  V. A. Andreev, D. M. Davidovi\'c, Lj. D. Davidovi\'c, M. D. Davidovi\'c, V. I. Man’ko
 and M. A. Man’ko,
Theoretical and Mathematical Physics, {\bf 166}(3): 356–368 (2011).

\bibitem{ADD1}  Andreev V.A., Davidovich D.M., Davidovich L.D., Davidovich M.D., {\it  Phys. Scr.}, {\bf 143}, (2011) 01400.

\bibitem{CG} K.E. Cahill and R.J. Glauber, {\it Ordered Expansions of the Boson Amplitude Operators},
Physical Review {\bf 177} No. 5 (1969) 1857.

\bibitem{BCH}
 E. Munguıa-Gonzalez,  S. Rego,  J. K. Freericks,
 {\it Making squeezed-coherent states concrete by determining their
 wavefunction},
Am. J. Phys. {\bf 89} (2021) 885–896.

\bibitem{AWSS} Alfred Wünsche, {\it Squeezed Coherent States in Non-Unitary
Approach and Relation to Sub- and
Super-Poissonian Statistics},
Advances in Pure Mathematics {\bf 7} (2017) 706-757.

\bibitem{Polj} Tomasz Linowski and Łukasz Rudnicki,
{\it Relating the Glauber-Sudarshan, Wigner and Husimi quasiprobability distributions
operationally through the quantum limited amplifier and attenuator channels},
 Phys. Rev. {\bf A 109} (2024) 023715.

\bibitem{BBC} J. S. Ben-Benjamin, L. Cohen,
{\it Quasi-distributions for arbitrary non-commuting operators},
Physics Letters A
Vol. {\bf 384}, Is.18, (2020) 126393.

\bibitem{AA} Arash Azizi, {\it Displaced Janus States: Tunable Non-Gaussianity and Exact Higher-Order Coherences for Quantum Advantage},
arXiv:2508.09234v1 [quant-ph].

\bibitem{MW} L. Mandel and E. Wolf, {\it "Optical Coherence and  Quantum Optics"}, Cambridge University Press (1995) 896 p.

\bibitem{AT}  G.S. Agarwal  and  K. Tara,  {\it Phys. Rev. A}  {\bf 47} (1993) 3160-3166.

\end{thebibliography}
\end{document}